\documentstyle[preprint,aps]{revtex}
\input epsf.tex
\def\DESepsf(#1 width #2){\epsfxsize=#2 \epsfbox{#1}}
\begin{document}

\preprint{\vbox{\hbox{OITS-634}}}
\draft
\title {Leptoquark explanation of HERA anomaly \\in the context of
gauge unification}
\author{ N.G. Deshpande and B. Dutta}
\address{Institute of Theoretical Science, University of Oregon, Eugene,
OR 97403}
\date{July, 1997}
\maketitle
\begin{abstract} 
We examine the consequences of leptoquark explanation of HERA anomaly
in the context of R parity conserving  supersymmetric gauge
unified theory with the gauge unification scale at $\sim 10^{16}$ GeV. We pointed
out the difficulty of constructing a grandunified theory. However gauge
unification is still possible at $\sim 10^{16}$ GeV when additional multiplets
are introduced. We determine the mass spectrum of these additional fields
(fermions and scalars) in gauge mediated and supergravity scenarios. Unique
signatures and mass bounds are discussed. 
\end{abstract}
\pacs{PACS numbers: }
\newpage Recent observation of high Q$^2$ anamoly in $e^+p$ scattering
seen by H1\cite{H1} and ZEUS\cite{ZEUS} collaboration at HERA, if
confirmed by further experimentation, would be a clear indication of
physics beyond the Standard Model (SM). Explanation of this anomaly have
been advanced in terms of contact interactions generated by physics
beyond the SM\cite{contact}, in terms of s-channel leptoquark
production\cite{lepto} or in terms of modification of the parton
distribution function\cite{tung}. Contact interactions have to satisfy
various constraints e.g. atomic parity violation, LEP data, CCFR data
etc. Although it is possible to construct such models, these typically
predict similar anomalous events in $e^-p$ data and excess
$e^+e^-$ events in Drell-Yan process at high Q$^2$ in FERMILAB, evidence
for either is lacking at present. Leptoquark explanation has been
pursued in the context of R-parity conserving or R-parity
violating supersymmetric theory. These states occur as
resonances at a mass of $\sim$ 200-220 GeV. In this paper we shall
pursue the leptoquark explanation of HERA data, but in the context of
R-parity conserving supersymmetric gauge unification. We prefer to
maintain R invariance because this allows the lightest SUSY particle to
be a candidate for dark matter. Further, in R violating scenarios \cite{gb} there
are numerous arbitrary coupling  e.g 27 of
$\lambda^{\prime}$($\lambda^{\prime}_{ijk}L_iQ_jD_K^c$), 9 of
$\lambda$($\lambda_{ijk}L_iL_jE_K^c$) and 9 of
$\lambda^{\prime\prime}$($\lambda^{\prime\prime}_{ijk}U^c_iD^c_jD_K^c$), where
$L_i$, $Q_i$ are the lepton and quark doublets and $E^c, U^c, D^c$ are the
lepton, uptype and down type singlets respectively; and also
$\lambda^{\prime}$ and $\lambda^{\prime\prime}$ can not appear together in
the same superpotential, since that gives rise to rapid proton decay. In
this present paper we shall demand that any additional leptoquark states
postulated in the
$SU(3)\times SU(2)\times U(1)$ context be part of supersymmetric gauge
unification picture so as to be theoretically attractive. 
Since the leptoquark mass is 200-220 GeV, we expect that there are other fields 
around this mass in order to maintain gauge unification.

We first consider gauge unification in the context of a grandunified theory
like SU(5) \cite{HR}. Lowest dimensional representation of SU(5) that we can have
leptoquarks in is the {\bf 10} dimensional representation. However, because
{\bf 10} dimensional representation is formed from the antisymmetric
combination of ${\bf \bar 5 \times \bar 5}$, its SU(5) invariant coupling
${\bf \bar 5_i\bar 5_j 10_k}$ is antisymmetric in the generation indices i and j.
Thus the leptoquark can not couple to the first generation quarks and lepton as
needed to explain the HERA events. One then has to consider a higher dimensional
representation like ${\bf 15}$, which is symmetric in the product ${\bf \bar
5\times
\bar 5}$. To remove gauge anomalies one also requires ${\bf{\bar {15}}}$
representation, which also makes the representation vector like. However
${\bf {15}}$ representation has diquark fields which belong to (6,1,-2/3)
representation of $SU(3)_C\times SU(2)_L\times U(1)_Y$. This along with the
leptoquark in (3,2,1/6) representation makes the couplings blow up as they
evolve  at around
$10^{11}$ GeV, long before unification. This is true for all representation with
higher dimensions. We have also investigated flipped $SU(5)\times U(1)$ models
\cite{su5u1} where leptoquarks in ${\bf{\bar{10}}}$ dimensional representation
can arise from a product of
$10\times e^c$. These states can give rise to leptoquarks in $e^-p$ scattering as
well as
$e^+p$. Leptoquarks of the type
$Qe^c$ and $d^ce^c$ are present. We will point out that their masses at the low
energy scale are very close to each other in any realistic scenario.  
In this paper we give up grandunification for the most part, and instead consider
the limited assumtion of gauge coupling unification at a large scale.  

 We shall assume that there is a leptoquark field with transformation (3,2,1/6)
and to remove anomalies, the conjugate representation is also
present. The leptoquark field will couple to $d_c$ and ($\nu, e^-$), but the
conjugate fields have no direct coupling to matter. The conjugate field has
gauge couplings that will allow them to be produced in $p\bar p$ collisions  at
FERMILAB, but only indirectly through mixing at HERA. Since there are terms in
the potential which  allows mixing between the field and the conjugate field, the
signature will bethe  same as the original field. Consequently, if these masses
are not well seperated, Tevatron could have seen or will see the effect of this
additional state in the near future. To maintain perturbative gauge unification
we now add diquark and dilepton content of {\bf 10} representation (${\bar
3}$,1,-2/3) and (1,1,1) and their conjugate. We shall refer to these multiplets
of new fields as
$f$ and their conjugate as
${\bar f}$. Another interesting aspect of the R-parity conserving theory is the
presence of the fermionic partners of the leptoquarks-the leptoquarkino \cite
{dmn}. The leptoquarkino decays into missing energy plus the same final state as
a leptoquark does. The leptoquarkino production crosssection at Tevatron is
higher than the leptoquarks for the same mass, since they are fermions.
Consequently the leptoquarkino has to be heavier than the leptoquark to escape
detection. This can be a problem since our natuaral expectation is the fermion
masses are smaller than the scalar masses. We will show that it is possible to
have the fermion masses become larger than the scalar masses by adding a singlet
field.  Since we are not using the SU(5), we can couple the leptoquark to just
the first generation, diquarks and dileptons need to couple to additional
generations. We will consider the possibility that the leptoquark couples to
other generation and derive bounds on the coupling to the matter fields. 
Our strategy is to write the superpotential involving the new fields and find the
mass spectrum of both the fermions and scalars at low scale using
 appropriate renormalization group equations (RGEs) and the D-terms
using two specific supersymmetric breaking models. We shall consider both the
gauge mediated SUSY breaking scenario as well as gravity induced SUSY breaking
since the process of transmission of information of SUSY breaking to the
observable sector is still an area of active interest. We shall explore the
experimental signature for the new fields in present and future colliders.

We first show that with the choice of multiplets gauge unification is
preserved even in two loop analysis. In Fig.1 we show the two loop evolution
of gauge couplngs assuming $\alpha=1/127.9$, ${\rm sin}^2\theta_W=0.2321$ and
$\alpha_3=0.118$ at $M_Z$. If either the diquark or dilepton is left out, the
gauge unification is not possibile.  As mentioned, a similar attempt with higher
dimensional representation will cause coupling to diverge well below
unification. In order to calculate the mass spectrum of this model we need a
superpotential. Let us first discuss the superpotential in the  gauge mediated
symmetry breaking (GMSB) scenario. The gauge mediated SUSY breaking models have
become very popular in last one year for several reasons. Since the SUSY
breaking is communicated to the obsevable sector by SM gauge group or some other
gauge group, the soft terms in these models are naturally flavor symmeteric.
These models also have less number of input parameters than the models where
SUSY breaking is communicated by gravity eg. A (coefficent of the trilinear
term)  term and B term (coefficent of the bilinear term) in the potentials are 0
at the GMSB sacle. The lightest supersymmetric particle (LSP) in these models is
gravitino and the next to LSP (NLSP) is either a stau or a neutralino, depending
on the parameter space. The signals of these kinds of SUSY breaking are
spectacular since  they have hard photon (neutralino is the NLSP) or a lot of
taus (stau is the NLSP) in addition to the missing energy  in the final states. 

The superpotential with the leptoquarks at the GMSB scale can be written
as:
\begin{eqnarray} W&=&\lambda_1\phi_{\nu e}{\bar
\phi_{\nu e}}S_1+\lambda_2\phi_{d^cd^c}{\bar
\phi_{d^cd^c}}S_1+\lambda_3\phi_{Ld^c}{\bar
\phi_{Ld^c}}S_1\\\nonumber 
&+&M_{Ld^c}\phi_{Ld^c}{\bar
\phi_{Ld^c}}+M_{d^cd^c}\phi_{d^cd^c}{\bar
\phi_{d^cd^c}}+M_{\nu e}\phi_{\nu e}{\bar \phi_{\nu e}}+{M^\prime\over
2}S_1S_1
\end{eqnarray}
where $S_1$ is a SM singlet, the necessity of the
singlet field will be discussed later.
$M_i$'s are the soft breaking masses that get introduced by the same
mechanism that generates the $\mu$  at the GMSB scale.  We will make an ansatz
that there is a common value for $M_i$ when we perform the calculation.
Similarly, to make the theory predictive, we assume the
$\lambda$s are unified to a comman value of 1. When we write the potential from
this superpotential there are terms like
$A_i\lambda_i\phi_i{\bar
\phi_i}S_1$ and
$M_iB_i\phi_i{\bar \phi_i}$. There are actually three more
terms in the superpotential involving the leptoquark and the quark and
lepton field, the diquark and the two quark fields and the dilepton and
two lepton fields These terms arise from from a GUT scale term like
$h_k L d^c\phi$. The coupling $h_k$required to
explain the HERA events is quite small ($\sim $0.05). Since the diquarks are
color triplet (antisymmetric combination of two color triplet fields) fields
and the dilepton are SU(2) singlet field, they need at least two generations for
this kind of coupling to survive. We can assume that these new field couple
to first and the 3rd generation.In fact if the leptoquarks $(Ld^c)$ are coupled
to the first generation only, the latest CDF bound rules out the
leptoquark  mass of about 210 GeV at $95 \% $ C.L.\cite{talk} using the
NLO QCD calculation \cite{kramer}. If we assume that the leptoquark couples to
all generations flavor diagonally, then
$K\rightarrow\mu e$ gets an unacceptable tree level contribution. We
therefore forbid the leptoquark coupling to the second generation.
However leptoquark can still couple to the third generation. This
induces tree level contribution to
$B\rightarrow\tau e$. The bound we get is
$h_1\times h_3<5.7\times 10^{-3}$, where
$h_1 $ and
$h_3 $ are the leptoquark coupling to the first and the third
generation. Consequently a third generation coupling to the leptoquark
of the same magnitude as the first generation ($\sim 0.05$) is allowed.
Furthermore the inclusion of the third family implies that the
leptoquark can decay into a $\tau$ and a b quark.In this case the
crossection needs to be multiplied by a factor of 1/4 and the leptoquark
mass bound reduces to 190 GeV.

There are also soft supersymmetry breaking gaugino
(gluino, wino, bino) and scalar masses(squarks, sleptons, Higgs, fields in $f$
and
$\bar f$) due the gauge mediated interaction with the messenger fields. These
masses at the messenger scale M are given by
\cite{{DN},{SPM}}
\begin{equation}
\tilde M_i(M) = n\,g\left({\Lambda\over M}\right)\,
{\alpha_i(M)\over4\pi}\,\Lambda.
\end{equation} and 
\begin{equation}
\tilde m^2(M) = 2 \,(n)\, f\left({\Lambda\over M}\right)\,
\sum_{i=1}^3\, k_i \, C_i\,
\biggl({\alpha_i(M)\over4\pi}\biggr)^2\,
\Lambda^2.
\end{equation} where $\alpha_i$, $i=1,2,3$ are the three SM gauge
couplings  and 
$k_i=1,1,3/5$ for SU(3), SU(2), and U(1), respectively. The $C_i$ are
zero for gauge singlets, and 4/3, 3/4, and $(Y/2)^2$ for the fundamental
representations of
$SU(3)$ and $SU(2)$ and
$U(1)_Y$ respectively. Here $n$
corresponds to number of 
$(5+{\bar 5})$ multiplets, and  $g(x)$ and
$f(x)$ are messenger scale threshold functions with $x=\Lambda/M$. The
net scalar mass of the $\phi_{Ld^c}$ and $\phi_{e\nu}$,$\phi_{d^cd^c}$
fields have contribution from $M_i$ and as well as $m_0$. The masses of the
fields in $f$ and $\bar f$ are same at the gauge mediated scale. We run these
$\lambda$'s from the GUT scale to the GMSB scale by the following RGEs :
\begin{eqnarray} 2{\cal D}\lambda_1 &=&
\lambda_1(-\sum_i{C1_i(4\pi\alpha_i)}+
5\lambda_1^2+3\lambda_2^2+6\lambda_3^2),\\\nonumber 2{\cal D}\lambda_2
&=&
\lambda_2(-\sum_i{C2_i(4\pi\alpha_i)}+
\lambda_1^2+5\lambda_2^2+6\lambda_3^2),\\ 2{\cal D}\lambda_3 &=&
\lambda_3(-\sum_i{C3_i(4\pi\alpha_i)}+
\lambda_1^2+3\lambda_2^2+8\lambda_3^2),\nonumber
\end{eqnarray} where  $i$ refers to the gauge group, 
${\cal D}\equiv{{16\pi^2}\over 2}{d\over{dt}}$ and 
\begin{eqnarray}\nonumber C1=\left( 0,0,{12\over 5}\right)\, ,\,
C2&=&\left({16\over3},0,{16\over 15}\right) , C3=\left(
{16\over3},3,{1\over 15}\right)\nonumber,
\end{eqnarray} We need to run the soft
masses, B terms, A terms from the GMSB scale down to the weak scale wfor
the new fields as well. The RGE's for the soft masses are given as: 
\begin{eqnarray} {\cal D}m_{j}^2 &=& -\sum_i{Cj_i(4\pi\alpha_i){\tilde
M_i}^2}+
\lambda_j^2(m_{j}^2+{\bar m_{j}}^2+{M_{S_1}^2\over 2}+A_{j}^2),
\end{eqnarray} where
j=1,2,3 represent  $Ld^c,d^cd^c,\nu e$. We also have a similar
equation for
$\bar m_j$. $\tilde
M_i$'s are the guagino masses. The A's evolve according to the following
RGE's:
\begin{eqnarray} {\cal D}A_{Ld^c} &=& \sum_i{C3_i(4\pi\alpha_i){\tilde
M_i}}+ 6\lambda_3^2A_{Ld^c}+5\lambda_2^2A_{d^cd^c}+\lambda_1^2A_{\nu
e},\\\nonumber {\cal D}A_{d^cd^c} &=& \sum_i{C2_i(4\pi\alpha_i){\tilde
M_i}}+ 8\lambda_3^2A_{Ld^c}+3\lambda_2^2A_{d^cd^c}+\lambda_1^2A_{\nu
e},\\ {\cal D}A_{\nu e} &=& \sum_i{C1_i(4\pi\alpha_i){\tilde M_i}}+
6\lambda_3^2A_{Ld^c}+3\lambda_2^2A_{d^cd^c}+3\lambda_1^2A_{\nu
e}\nonumber
\end{eqnarray}
Similarly, B's evolve according to the following RGE's:
\begin{eqnarray} {\cal D}B_{j} &=& \sum_i{Cj_i(4\pi\alpha_i){\tilde
M_i}}+ 2\lambda_j^2A_{j}
\end{eqnarray}
The mass parameters M's evolve according to the RGEs:
\begin{eqnarray} 2{\cal D}M_{j} &=& M_{j}(-\sum_i{Cj_i(4\pi\alpha_i)}+
2\lambda_j^2)
\end{eqnarray}
We can similarly write down the RGE for the soft 
breaking mass of the singlet $S_1$. We use the ref.\cite{vernon} for the
RGE's for the gauge couplings,for the other soft masses e.g the gaugino
($\tilde M_i$), squark and slepton masses and for the other parameters
like
$A_{t,b,\tau}$.  We keep
$\mu$ as free parameter which is determined at the weak scale  in the
tree level by:
\begin{eqnarray} {M_Z^2\over
2}&=&{{M_{H_1}^2-M_{H_2}^2tan^2\beta}\over{tan^2\beta-1}}-\mu^2
\end{eqnarray}  where $\tan\beta={v_2\over v_1}$.
 After we run these
equations down to the weak scale we have to include the D term
contribution to the scalar masses, which  is of the form:
\begin{eqnarray}
m^2_D(T_3,{Y\over2})&=&\pi(v_1^2-v_2^2)(\alpha_2T_3-\alpha_1{Y\over2}{3\over5})
\end{eqnarray} These D-terms cause a mass splitting between $f$
and $\bar f$ scalars. Further, there is mixing between $f$ and ${\bar
f}$ scalars through the B term in the potential. This is a
large effect for diquarks and leptoquarks but small for dileptons. This
is because B term can be large for color fields at the weak scale even if
its  boundary value is 0 at the GMSB scale. The actual mixing for the
leptoquark is very nearly 45$^0$. Consequently at HERA, the lighter
combination of the mixed state is produced although
$e^+q$ couples only to the f representation. The higher mass
state is out of reach for HERA with the present energy. Similarly at
Tevatron, one can produce either state through gluon exchange, but mass
limits apply to to the lower mass state. The mass of the singlet field
and the coupling of the singlet to the leptoquarks are needed here since
if we look at RGE's we find the gauge part and the Yukawa part act in the
opposite direction and we will want the leptoquark mass to be lower than
the leptoquarkino mass.   This is because the leptoquarkino production
crosssection is always larger than the leptoquark
production cross sections for the same mass, and the final state in the
case of a leptoquarkino pair production will also have the leptons
and jets. So the same bound which applies to leptoquark will also apply
to the leptoquarkino. 
In Table 1  we show the masses of the new fields as well as the masses
of the  squarks, sleptons, neutralinos, charginos and gluino. The
input parameters at the GMSB scale are $\Lambda$, $M$, n, $\tan\beta$,
$M_i$s and $M^\prime$. The Yukawa couplings for the regular fermions are
evaluated from their masses. The Yukawa coupling ($\lambda_i$) for the
leptoquarks are assumed to be unified at the GUT scale and its value at
that scale is taken to be 1. At
the GMSB scale 
$\lambda_1=0.3$,
$\lambda_2=0.7$,$\lambda_3=0.8$. We have assumed the masses $M_i$
(where i=$Ld^c,d^cd^c,$and $\nu e$) to be the same at the GMSB scale. We vary
the common mass
$M$ and $M^\prime$ in such a way so as to get the leptoquark mass around
200-220 GeV. This fixes the diquark and dilepton masses also. From the
table 1 it is easy to see that the leptoquark mass
$\sim 200-220$ GeV can be obtained over a  wide range of parameter space.
In that parameter space the diquarks are much heavier than the
leptoquarks and the leptoquarkinos are also heavier than the
leptoquarks. The dileptons can be  lighter or heavier. We can see from
the table that the mass states of
$\phi_{Ld^c}$ are well seperated due to the large B term.  This is also
true for the diquarks  but not for the dileptons where  $B$ is small.
If we take the $M_i$'s in the ratio of the Yukawa coupling
($\lambda_i$'s), the mass spectrum of the scalars would remain
unchanged, only the dileptino mass will reduce to almost half the value
showed in the tables.
  
Let us discuss the signals of the new fields in different
colliders. In Tevatron the diquarks will be produced pairwise through s
channel gluon exchange just like the squarks. Each diquark decays into a
pair of jets. So we have 4 jets in the final state. The dileptons 
can be seen at the Tevatron through Drell-Yan production with subsequent
decay into
2 leptons plus missing energy. Similar signals can be produced
by the charged Higgs pair, W pair or the selectron pair. Leptoquark
pair decays into 2jet+2lepton. Beside Tevatron these particles can be
pair produced in the electron-positron collider.  One can also produce
leptoquark
\cite{don} and dilepton singly in e$\gamma$ collider through a t channel
exchange. The final state will be
$l$+jet in the case of leptoquark and  a 2 leptons  plus missing
energy (one lepton has to be other than electron, since the dileptons are
connected to two leptons of different generations) in the case of dilepton. The
fermionic partners of these fields:  leptoquarkino, dileptino or diquarkino have
interesting signatures in the colliders. The diquarkino will decay into a quark
and a squark. The squark will eventually decay into a quark and neutralino (quark
gluino channel is not suitable since gluino mass is large in GMSB models). The
neutralino will give rise to a photon and a gravitino. When diquarkinos are pair
produced the final state wil have 4 j+2 hard photons +missing energy. In the
case of stau being the NLSP the diquarkino will decay into 4j+4$\tau$ lepton
+missing energy, since each neutralino will deacy 100$\%$ of the time into
2$\tau$ +plus missing energy. The
$P_T$ distribution of those 4
$\tau$s will be different and two of them will have very high $P_T$. This
signature is very hard to miss. The production of these
diquarkinos will be similar to that of top quark. The
diquarkino mass of about 370 GeV will approximately give rise to  one
event  ($4 jets+4\tau $or$ 2\gamma$ plus missing energy) in Tevatron
with a luminosity of 110
$pb^{-1}$. So far 6 events having 4 jets plus missing energy (missing
energy$>60$ GeV have been observed \cite {cdf2}, but they are claimed to
be consistent with the SM precesses plus detector induced background. The
dileptino
 will decay to a lepton and a
sneutrino or neutrino and slepton and sneutrino or slepton will
decay into neutrino or lepton and a neutralino, which will give rise to
final states 
lepton+missingenergy (slepton mass is less than the neutralino
mass) when stau is lightest or the 
lepton+$\gamma$+ missing energy when neutralino is the lightest in the
gauge mediated models. Consequently when they are produced in Tevatron
or electron positron collider the dileptino pair can give rise to
$l^+l^-+2\gamma$+missingenergy signal. One such
event 9where both the leptons are electrons) was claimed to be observed 
recently\cite{SP}. The same signal can be produced by the selectrons also. In an
e$\gamma$ collider the dileptino can be singly produced in the association of
sneutrino through t-channel exchange of dileptino. Sneutrino decay gives
rise to $\gamma$ plus missing energy (neutralino is NLSP) or 4$\tau$ plus missing
energy (lighter stau is the NLSP) in the final state. The  leptoquarkino decays
into a spositron (spositron decays into a positron and a neutralino with
$\chi_0\rightarrow\gamma\tilde G$) and d quark (or squark and an electron, but
the squark masses are large in these models) with the final state in the
leptoquarkino pair production process has 
$e^+e^-\gamma\gamma$jj plus missing energy without any SM
background.  If lighter stau is the NLSP and neutralino is the NNLSP, 
the  final state in the leptoquarkino pair production process has 
$e^+e^-2\tau^+2\tau^-$jj plus missing energy (since in this
case$\chi_0\rightarrow 4\tau$ plus missing energy) . Out of these six leptons in
the final state, one
$\tau$ pair (produced from the decay of stau) has much higher $P_T$ than
the other leptons. So far each CDF\cite{cdf} and D0\cite{D0} have
reported one event in the  $e^+e^-$jj plus missing energy channel. If we
apply that bound we get leptoquarkino mass has to be greater than 345
GeV, since the pair producton crosssection is around 0.04 pb and $e^+e^-$
detection efficiency is 0.2 assuming the leptoquarkino branching ratio
to be 1. If however leptoquarkino branching ratio is 1/2 the bound
becomes 290 GeV. One could also look for the signals of
leptoquarkino production in
$e^+\gamma$ colliders. The leptoquarkino can be singly
produced in such a  collider.  In the gauge mediated SUSY
breaking scenario, the final state has either a hard photon or
$\tau^+\tau^-$ along with electron +jets+missing energy. In all the
above cases the fermion component of the new multiplet can be
distinguished from the spin 0 component by hard photon or excess $\tau$s
and the missing energy.

If we had used the leptoquarks in SU(5)$\times U(1)$ unifying group, we would
have 3 leptoquarks and a dilepton as discussed before. In this case the above
analysis would be unchanged. However all the leptoquarks will have almost the
same soft symmetry breaking mass at the GMSB scale (since all of them have
color quantum number) and their masses at the weak scale also will be very close
(D term can  not produce much of a splitting). This means that not only HERA,
but also the Tevatron will see these 3 leptoquarks within a small mass
range. The dilepton ($\nu^c e^c$) can be produced at FERMILAB through
Drell-Yan mechanism and has an interesting signature ($e^+e^-e^+e^-jjjj$)
arising from the decay of dilepton into $e\nu_R$, followed by
$\nu_R\rightarrow eW^-\rightarrow ejj$.

 In the supergravity motivated theory the soft masses are introduced at
the GUT scale. We assume the squark, Higgs and slepton the new f
multiplet all have same soft mass
$m_0$. The gaugino masses are also unified to a common mass
$m_{1/2}$. The other parameters are the bilinear coefficient  of the
$f{\bar f}$ and the
$S_1S_1$term in the superpotential and  the coefficient of the same
bilnear terms in potential. Unlike the gauge mediated model the bilinear
coefficent need not be  0 to start with.  In the case of supergravity motivated
theory the signals are different from the GMSB models signals. Since neutralino
is the LSP, there is no hard photon or high $P_T$ $\tau$s in the final
state. For example, the diquarkino would decay into quark and squark
and the squark will deacy into a quark and missing energy. The final
state will be 4jets+missing energy when the diquarkino is pair produced.
The leptoquarkino pair will give rise to a jjl$^+$l$^{-}$ plus missing
energy and the dileptino pair will give rise to l$^+$l$^{-}$ plus
missing. The missing energy
however would distinguish the fermion signal from the boson one.

In Table 2  we show the masses of the new fields as well as the masses
of the  squarkss, sleptons, neutralinos, charginos and gluino. The
input parameters at the GUT scale are the universal scalar mass $m_0$,
universal gaugino mass $m_{1/2}$, $M_i$'s, $M^\prime$, $\tan\beta$, $A$'s
and $B$'s. Again for simplicity  we have assumed the masses $M_i$ to be
same at the GUT scale. We also have assumed that other than the $B$
associated with the f and ${\bar{f}}$, all the
$B$'s are 0  at the unification scale. We also assume that all the $A$'s
are 0 at the GUT scale. From the table 2 it is easy to see that the
leptoquark mass
$\sim 200-220$ GeV can be obtained in  wide range of parameter space. In
that parameter space the diquarks are much heavier than the leptoquarks
and the leptoquarkinos are also heavier than the leptoquarks. The
dileptons can be  lighter or heavier. The lighter combination of
$\phi_{Ld^c}$ and
${\bar {\phi_{Ld^c}}}$ is again much lighter than the heavier combination

So far we have considered the case with just one type of leptoquark
i.e $\phi_{Ld^c}$. If the charge current anamoly is also established, 
which is not at all clear from the present data, we shall need an
additional representation like {\bf 45} which has $Lu^c$ type of
leptoquarks. These leptoquarks will then mix because of symmetry
breaking with the $Ld^c$ type of leptoquarks and using the technique
showed in the ref. \cite{babu} one can get a charge current excess. If
charged current events do hold up, such additional multiplets will need
to be considered.

In conclusion we have used leptoquark in the context of gauge unification and
have  discussed  the mass spectrum and the signatures of the additionals fields
needed to maintain the unification. We have  showed that the fermionic
partner of the leptoquark will be heavier than the scalar part. We have also
showed that one combination of the leptoquark field and  conjugate field
(we need to add in order to cancel the anomaly) is lighter than the other
combination, even though the soft mass generation mechanism is same for
both types of fields. The heavier combination is still out of reach of
Tevatron. 

We would like to thank K. S. Babu and S. Nandi for useful discussions. The work 
has been supported by a DOE grant no. DE-FG06-854ER-40224.

\newpage
\begin{center} {\bf TABLE CAPTIONS}\end{center}
\begin{itemize}
\item[Table 1~:] {Mass spectrum for the new fields (diquarks,
leptoquarks and dileptons), stop and sbottom squarks (m$_{\tilde {\rm
t}}$,m$_{\tilde {\rm
b}}$), stau lepton(m$_{\tilde
{\tau}}$), chargino ($\chi^{\pm}$),
lightest neutralino (m$_{\chi^0}$), $\mu $, gluino mass (m$_{\tilde
g}$) are shown in the gauge mediated supersymmetry breaking model. The
scalar component of the new fields are written as m$_i$ and the fermionic
components are written as M$_{i}$}. The two entries for the scalars
give the lighter and the heavier mass eigenstates.
\item[Table 2~:] {Same mass spectrum as  in Table 1 is shown for the
supergravity scenario}.
\end{itemize}
\begin{center} {\bf FIGURE CAPTIONS}\end{center}
\begin{itemize}
\item[Figure 1~:] {The evolution of the coupling constants are shown in
presence of the new fields using two loop RGEs. }.
\end{itemize}

\newpage
\begin{center}  Table 1 \end{center}
\begin{center}
\begin{tabular}{|c|c|c|c|c|c|}  \hline
&Scenario1&Scenario2&Scenario3&Scenario4&Scenario5\\\hline
masses&$\Lambda=60$ TeV,&$\Lambda=60$TeV,&$\Lambda=60$ TeV,&$\Lambda=30$
TeV,&$\Lambda=25$TeV,\\
(GeV)&n=1, $M=4\Lambda$&n=1, $M=4\Lambda$&n=1, $M=10^2\Lambda$&n=2,
$M=10^2\Lambda$&n=2,
$M=10^3\Lambda$\\
&$\tan\beta$=3&$\tan\beta$=35&$\tan\beta$=3&$\tan\beta$=3&$\tan\beta$=12
\\\hline m$_{\phi_{d^cd^c}}$&371,580&372,581&346,630&307,596&271,579\\\hline
M$_{{d^cd^c}}$&388&388&427&418&444\\\hline
m$_{\phi_{e\nu}}$&152,174&151,177&207,225&244,259&270,287\\\hline
M$_{{e\nu}}$&305&305&309&309&311\\\hline
m$_{\phi_{\nu d^c}}$&210,500&210,510&210,592&210,570&203,570\\\hline
m$_{\phi_{ed^c}}$&210,500&210,510&210,592&210,570&203,570\\\hline
M$_{Ld^c}$&399&400&444&434&463\\\hline
m$_{\chi^0}$&78&80&77&75&72\\\hline
m$_{\chi^{\pm}}$&143,403&148,347&144,446&135,346&113,274\\\hline m$_{\tilde
{\tau}}$&111,222&54,223&119,230&89,171&65\\\hline  m$_{\tilde {\rm
t}}$&614,719&633,705&542,678&441,564&373,475\\\hline m$_{\tilde {\rm
b}}$&685,726&653,700&638,655&509,547&414,435\\\hline m$_{\tilde
g}$&538&538&534&535&447\\\hline
$\mu$&-379&-322&-426&-315&-238\\\hline
\end{tabular}
\end{center}

\newpage
\begin{center}  Table 2 \end{center}
\begin{center}
\begin{tabular}{|c|c|c|c|c|c|}  \hline
&Scenario1&Scenario2&Scenario3&Scenario4&Scenario5\\\hline
masses&$
m_0=250$GeV,&$m_0=265$GeV,&$m_0=280$GeV,&$m_0=320$GeV,&$m_0=230$GeV,\\
(GeV)&$m_{1/2}=250$GeV,&$m_{1/2}$=200GeV,&$m_{1/2}=180$GeV,&$m_{1/2}=160$GeV,&
$m_{1/2}=240$GeV\\
&$\tan\beta$=3&$\tan\beta$=5&$\tan\beta$=30&$\tan\beta$=10&$\tan\beta$=10
\\\hline
m$_{\phi_{d^cd^c}}$&350,789&321,651&304,605&302,562&333,727\\\hline
M$_{{d^cd^c}}$&455&401&401&402&566\\\hline
m$_{\phi_{e\nu}}$&183,306&202,312&216,324&247,348&119,261\\\hline
M$_{{e\nu}}$&204&180&180&180&202\\\hline m$_{\phi_{\nu
d^c}}$&212,841&212,692&205,642&204,593&207,772\\\hline
m$_{\phi_{ed^c}}$&212,841&218,692&205,642&204,593&207,772\\\hline
M$_{Ld^c}$&512&401&451&452&565\\\hline
m$_{\chi^0}$&100&80&72&64&97\\\hline
m$_{\chi^{\pm}}$&191,502&148,389&136,338&322,345&185,428\\\hline
m$_{\tilde {\tau}}$&269,311&277,305&248,316&89,171&244,293\\\hline 
m$_{\tilde {\rm t}}$&443,671&372,567&354,513&329,501&446,640\\\hline
m$_{\tilde {\rm b}}$&596,650&498,548&429,498&438,498&570,619\\\hline
m$_{\tilde g}$&683&547&491&437&656\\\hline
$\mu$&-483&-477&-314&-301&-407\\\hline
\end{tabular}
\end{center}

\begin{figure}[htb]
\vspace{1 cm}

\centerline{ \DESepsf(uni.epsf width 9 cm) }
\end{figure}


\newpage
\begin{thebibliography}{[001]}

\bibitem{H1} C. Adloff et al., H1 collaboration, DESY 97-024, hep-ex/9702012.

\bibitem{ZEUS} J. Breitweg et al., ZEUS collaboration, DESY 97-025,
hep-ex/9702015.
\bibitem{contact} G.Altarelli, J. Ellis, G. Giudice, S. Lola, M.
Mangano, hep-ph/9703276; K. S. Babu, C. Kolda, J. March-Russell and F.
Wilczek, hep-ph/9703299; M. C. Gozales-Garcia and M. Fabbrichesi,
hep-ph/9703346; A. Nelson, Phys. Rev. Lett. {\bf 78}, 4159,(1997); V.
Barger et al., hep-ph/9703311; N. Di Bartolomeo and M. Fabbrichesi,
hep-ph/9703375; A. Nelson, hep-ph/9703379; W. Buchmuller and D. Wyler,
hep-ph/9704317; K. Akama, K. Katsuura and H. Terazawa, hep-ph/9704327.
N.G. Deshpande, B. Dutta and X.-G. He, hep-ph/9705236(to
appear in Phys. Lett.B)
\bibitem{lepto}
D.~Choudhury and S.~Raychaudhuri, hep-ph/9702392; H.~Dreiner and
P.~Morawitz,  hep-ph/9703279;
	M.~Doncheski and S.~Godfrey,
hep-ph/9703285;J.~Bl\"{u}mlein, hep-ph/9703287;
	J.~Kalinowski etal,hep-ph/9703288;G. Altarelli et. al. in ref.1;
K.S.~Babu etal in ref. 1;. M. Drees, hep-ph/9703332;
	J.L. Hewett, T. G. Rizzo, hep-ph/9703337]; G.K. Leontaris and J.D.
Vergados, hep-ph/9703338;Z.~Kunszt and W.~Stirling, hep-ph/9703427;
	T. Kon, and T. Kobayashi, hep-ph/9704221; R. Barbieri, A. Strumia, and
Z. Berezhiani hep-ph/9704275; I. Montvay, hep-ph/9704280];
	M. Kramer, T. Plehn, M. Spira, and P.M. Zerwas, hep-ph/9704322];
	G.F. Giudice and R. Rattazzi, hep-ph/9704339];
	S.F. King and G.K. Leontaris, hep-ph/9704336;
	A.S. Belyaev and A.V. Gladyshev, hep-ph/9704343; 
	B. Dutta, R.N. Mohapatra, and S. Nandi, hep-ph/9704428; G. Altarelli,
G.F. Giudice, and M.L.~Mangano, hep-ph/9705287. K.S.~Babu, C.~Kolda, and J.~March-Russell,
hep-ph/9705414;  S. Jadach, W. Placzek, B.F.L. Ward, hep-ph/9705395;
J. Ellis, S. Lola, K. Sridhar, hep-ph/9705416;
F.  Caravaglios,  hep-ph/9706288;
L. Giusti, A. Strumia, hep-ph/9706298;
Jihn E. Kim, P. Ko, hep-ph/9706387;
A.S. Joshipura, V. Ravindran, S.K. Vempati, hep-ph/9706482;
S. Lola, hep-ph/9706519; Z. Cao, X-G. He, B. McKellar, hep-ph/9707227;
E. Keith, E. Ma, hep-ph/9707214. 

\bibitem{tung} 	S. Kuhlmann, H.L.~Lai, and W.K.~Tung,
hep-ph/9704338; K.S.~Babu, C.~Kolda, and J.~March-Russell,
hep-ph/9705399.

\bibitem{gb} G. Bhattacharyya, Nucl. Phys. Proc. Suppl. {\bf 52A}, 83,1997 and
references therein.
 
\bibitem{HR} J. L. Hewett et al in ref.4 

\bibitem{su5u1} I. Antoniadis, J. Ellis, J.S. Hagelin and D.V. Nanopoulos,
Phys. Lett. {\bf B 194}, 321 (1987).

\bibitem{dmn} B. Dutta et. al in ref.4

\bibitem{DN} M. Dine and A. Nelson, Phy. Rev. {\bf D47}, 1277 (1993); M. Dine,
A. Nelson and Y. Shirman, Phys. Rev. {\bf D51}, 1362 (1995); M. Dine, A. Nelson,
Y. Nir and Y. Shirman, Phys. Rev. {\bf D53}, 2658 (1996); M. Dine, Y. Nir and Y.
Shirman, preprint SCIPP-96-30, hep-ph/9607397.

\bibitem{SPM} S. Dimopoulos, G.F. Giudice and A. Pomarol, preprint
CERN-TH/96-171, hep-ph/9607225; S. P. Martin, hep-ph/9608224.

\bibitem{vernon}V. Barger, M. Berger, P. Ohmann, and R. J. N. Phillips,
Phys. Rev.{\bf D51}, 2438 (1995), and references therein.

\bibitem{talk} C. Grosso-Pilcher (CDF-collaboration), talk at the Vanderbuilt
meeting, May 15th,1997.

\bibitem{kramer}M. Kramer et. al. in Ref. 2.

\bibitem{don} M. Doncheski, S. Godfrey, Talk given at 1996 DPF / DPB Summer
Study on New Directions for High-Energy Physics (Snowmass 96), Snowmass, CO, 25
Jun - 12 Jul 1996, hep-ph/9612385 and references therein. 

\bibitem{cdf2} F. Abe et al.,  The CDF Collaboration,
FERMILAB-PUB-97/031-E.

\bibitem{SP} S. Park, representing the CDF collaboration, ``Search for New
Phenomena in CDF," in {\it 10th Topical Workshop on Proton-Antiproton Collider
Physics}, ed. by R. Raja and J. Yoh (AIP Press, New York, 1995), report
FERMILAB-CONFE-95/155-E.

\bibitem{cdf} F. Abe et al., Phys. Rev. Lett. {\bf 74}, 2626 (1995); for
a recent summary, see D. S. Kestelbaum, FERMILAB-CONF-97/ 016-E.

\bibitem{D0} S. Abachi et al., Phys. Rev. Lett. {\bf 74}, 2632 (1995); 

\bibitem{babu} K. S. ~Babu et.al in ref.2;
\end{thebibliography}
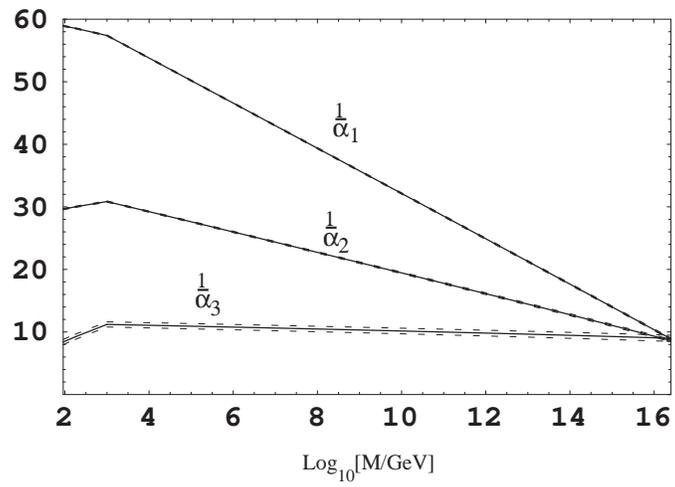
\end{document}